\documentclass[prd,reprint,showpacs,showkeys,superscriptaddress]{revtex4}
\usepackage{amsfonts}
\usepackage{amssymb}
\usepackage{amsmath}
\usepackage{graphicx}
\usepackage[font={footnotesize,it}]{caption}
\usepackage{latexsym}

\begin{document}

\title{Gravitational Lensing in Rotating and Twisting Universes}
\author{O. Gurtug}
\email{ozaygurtug@maltepe.edu.tr}
\affiliation{T. C. Maltepe University, Faculty of Engineering and Natural Sciences,
34857, Istanbul -Turkey}
\author{M. Mangut}
\email{mert.mangut@emu.edu.tr}
\affiliation{Department of Physics, Faculty of Arts and Sciences, Eastern Mediterranean
University, Famagusta, North Cyprus via Mersin 10, Turkey}
\author{M. Halilsoy}
\email{mustafa.halilsoy@emu.edu.tr}
\affiliation{Department of Physics, Faculty of Arts and Sciences, Eastern Mediterranean
University, Famagusta, North Cyprus via Mersin 10, Turkey}

\begin{abstract}
Gravitational lensing caused by the gravitational field of massive objects has been  studied and acknowledged for a long period of time. In this paper, however, we propose a different mechanism where the bending of light stems from the non-linear interaction of gravitational, electromagnetic and axion waves that creates the high curvature zone in the space-time fabric. The striking distinction in the present study is that in contrast to the convex lensing in the gravitational field of a massive object, hyperbolic nature of the high curvature zone of the background space-time may give rise to concave lensing. Expectedly detection of this kind of lensing becomes possible through satellite detectors.

\end{abstract}

\maketitle

\section{Introduction}

The experimental observation of gravitational waves has opened a new window to understand the universe in a more global scale. From the first observation to date, it is understood that our universe contains propagating gravitational waves. According to the Einstein's theory of general relativity, these gravitational waves, alone or coupled with electromagnetic (em) waves, do not pass through each other without a significant interaction. Their interaction is nonlinear and hence, induces irreversible consequences on the structure of the fabric of space-time. The remarkable influence on the space-time structure is the high curvature zones. \\

In our earlier studies, we have shown that nonlinear interaction of plane em shock waves accompanied with gravitational waves with different amplitude profiles generates  cosmological constant that can be associated with dark energy \cite{1}, which is believed to be the source of the accelerating expansion of our universe. In an another study, we have shown that the nonlinear interaction of plane gravitational waves and shock em waves with second polarization induces Faraday rotation in the polarization vector of em waves which paves the way for an indirect evidence of the gravitational waves \cite{2}. In this paper, we present the gravitational lensing in the high curvature zones which stem from the nonlinear interaction of gravitational waves coupled with em and axion waves such that any massive or massless particle that moves in such a region might be diffused.   \\

Bending of light while passing near a massive object such as a star or black hole is a well-known, one century old problem named as gravitational lensing. This is due to the fact that light has energy which is equivalent to mass and is attracted by another mass as a requirement of gravity. In particular, when light passes near a highly massive star, the deflection angle becomes significant enough to be detected by telescopes. \\

The purpose of this study is to introduce another mechanism rather than the gravitational field of a massive object that leads to gravitational lensing. This mechanism incorporates with the high curvature zone of space-time. In such space-times there is no definite center, thus any chosen point can act for the purpose and all deflections are computed with respect to that center. The nonlinear interaction ( or collision) of gravitational, em  and axion waves produces high curvature regions that are isometrically transformable to the spherical coordinates in which the bending angle calculations become more tractable. \\

Our first example in this direction is the rotating Bertotti-Robinson (RBR) space-time \cite{3,4},  which is isometric to the collision of two cross polarized em waves \cite{5,6}. It has been known that the collision of two linearly polarized em waves, the so-called Bell-Szekeres (BS) solution \cite{7}, can be transformed into the Bertotti-Robinson (BR) space-time \cite{8} which is isotropic and conformally flat (CF), so that light remains undeflected. However, the nonlinear interaction of the cross polarized em waves is locally isometric to the RBR space-time. The cross polarized nature of the waves that participate in the collision breaks the isotropy and induces a cross term in the metric and naturally deflects the light. In other words, the cross term amounts to a non-zero Weyl curvature which behaves like a mass term. The reason that we take collision of em waves into account can be summarized as follows: If we follow the Big Bang at the end of a period of nearly $400.000$ years, em interaction was switched on to allow strong collision of such waves to shape the future of the newly born universe. Further expansion naturally cooled down the em radiation to form the present patterns of cosmic microwave background (CMB) radiation. The energy density of em radiation which is described by the Newman-Penrose (NP) \cite{9} quantity $\Phi_{11}$ is dependent on the angle that can be plotted to exhibit the non-isotropy as a function of an angular variable. In particular, via an observation by a differential interval of very small angle, such as $0<\Delta \theta <1$ in degrees, the energy distribution can be plotted to show the variation. Such a distribution may be compared with the energy density of the CMB radiation. Note, however, that the general distribution depends on more factors, whereas in our present model we take only the rotational effects into account to distort the isometry. \\

Our second example consists of the space-time formed by the collision of  em waves coupled to an axion field \cite{10}. The axion also induces a cross term in the metric to bend the light. Applying a coordinate transformation to the em-axion problem, we obtain once more a non-isotropic space-time and we find in such a space-time the bending angle of light. It is interesting that light bending in this problem is provided entirely by the axion field. Vanishing of the axion field reduces the space-time once more to the isotropic BR in which the light shows no deflection. This particular example of axion is also important in the sense that axion is considered to contribute to the dark matter. Looking to the problem from the other direction, detection of deflection of light in such a space-time may be helpful in detection of the axion. \\

The last example that we shall consider within this context consists of the Newman-Unti-Tamburino (NUT) space-time \cite{11}. In a previous study, the NUT parameter was interpreted as the twist of the empty space-time \cite{12}. As the off-diagonal metric component gave rise to light deflection, the twist property of space-time also creates bending of light in the empty space. Let us note that apart from the twist interpretation, the NUT parameter can be interpreted as a gravito magnetic mass \cite{13}, as an extension of the Schwarzschild mass. Naturally, the latter also leads to the effect of empty space bending due to the curvature of the space-time. Expectedly, as the NUT parameter vanishes, one recovers the result of light bending nearby a Schwarzschild black hole, a well-known result. \\

One important distinction in the present study is that although masses are always attractive and give rise to convex lensing, in our examples, the lensing can also be concave. That is, the null geodesics are deflected in the outward direction by the rotation/twist that induces hyperbolic property to the space-time. For instance, in the case of the NUT space-time, if the NUT parameter is smaller / larger than the angular momentum parameter, then accordingly we encounter with a convex / concave type bending for the light trajectory. Similarly, the rotation parameter of BR  may shift the character of light bending due to the presence of axion. It is worth to emphasize at this stage that some researchers interrelated the concave lensing effect with the presence of dark energy \cite{14,15}. \\

Our method of finding light bending is based on the method developed by Rindler and Ishak  \cite{16}, which has been employed previously in different examples \cite{17,18,19,20}. In section II, we give a brief review of this formalism. In analogy with the Kepler's problem of Newtonian mechanics, we project the orbits onto the $\theta=\pi /2$ plane and express the $u=\frac{1}{r}$ as a function of azimuthal angle $\varphi$. Our case is just for the light orbits, as null geodesics which can be developed perturbatively in terms of $\varphi$. \\

Organization of the paper is as follows. In Section II, we review the general formalism of bending angle calculation in rotating/twisting geometries. Specific examples are given in Sec. III, such as rotating BR, axionic BR and twisting NUT spacetimes. We complete the paper with results and discussions in Section IV.

\section{General Formalism of the Bending Angle of Light for Rotating Geometries}

Among the others, the method proposed by Rindler and Ishak (RI) \cite{15} for calculating the bending angle of light has been found more powerful, especially,  when one wishes to display the effect of the background fields to the gravitational lensing. In this section, we extend this method to cover the metrics that describe rotating fields. The form of the metric that we will be interested in for the rotating geometries is defined at the equatorial plane as

\begin{equation}
ds^{2}=f(r)dt^{2}+2g(r)dtd\varphi-h(r)dr^{2}-p(r)d\varphi^{2}.
\end{equation}   \\
The method of RI incorporates with the inner product of two vectors that remains invariant under the rotation of coordinate systems. As a result, the angle between two coordinate directions $d$ and $\delta $, as shown in Fig. 1,
is given by the invariant formula

\begin{equation}
\cos \left( \psi \right) =\frac{d^{i}\delta _{i}}{\sqrt{\left(
d^{i}d_{i}\right) \left( \delta ^{j}\delta _{j}\right) }}=\frac{%
g_{ij}d^{i}\delta ^{j}}{\sqrt{\left( g_{ij}d^{i}d^{j}\right) \left(
g_{kl}\delta ^{k}\delta ^{l}\right) }}.
\end{equation}

\begin{figure}[htp]
\par
\label{figure}
\par
\begin{tabular}{cc}
\includegraphics[width=100mm]{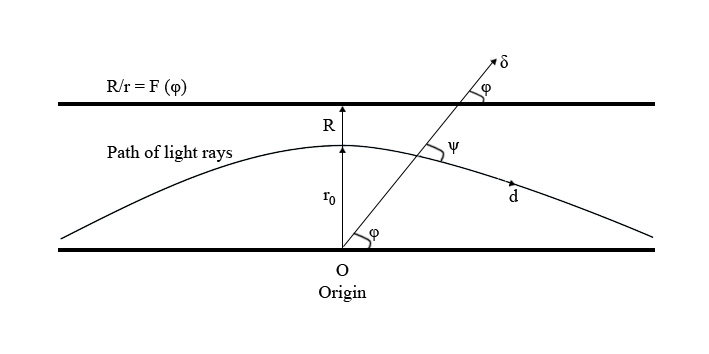} &
\end{tabular}
\caption{The orbital map of the light rays corresponding to Eq.(6). The one-sided bending angle is defined from this map as $\epsilon =\psi
-\varphi .$ The upper thick straight line represents the undeflected light rays described by the solution of the homogeneous part of the Eq.(6). The function $F(\varphi)$ is found to be $e^{\omega \varphi }$ and $e^{ \varphi }$ for the RBR and em-axion case respectively. In the case of twisting NUT universe, the function $F(\varphi)$ for small NUT parameter is $sin(\delta\varphi)$ and for large NUT parameter it reads $e^{\zeta \varphi }$.}\centering
\end{figure}

In this formula, $g_{ij}$ stands for the metric tensor of the constant time
slice of the metric (1). In this method, the orbital plane of the light rays is defined by introducing a two-dimensional curved $(r,\varphi )$ space, which is defined simultaneously at the equatorial plane (when $\theta =\pi /2$ ) and a constant time slice,
\begin{equation}
dl^{2}=h(r)dr^{2}+p(r)d\varphi ^{2}.
\end{equation}%
The constants of motion related to the null geodesics in the considered
rotating spacetime are%
\begin{equation}
\frac{dt}{d\lambda }=\left[ \frac{-p(r)}{K(r)} \right]E+ \left[ \frac{g(r)}{K(r)} \right] l,\text{ \ \ \
\ \ }\frac{d\varphi }{d\lambda }= \frac{-g(r)E-f(r)l}{K(r)},
\end{equation}%
in which $\lambda $ stands for the parameter other than  proper time and $K(r)=g^{2}(r)+p(r)f(r)$. From these conserved quantities,
we obtain%
\begin{equation}
\left( \frac{dr}{d\varphi }\right) ^{2}=\frac{K(r)}{h(r)\left[g(r)E+f(r)l)\right]^{2}}\left[p(r)E^{2}-f(r)l^{2}-2g(r)lE\right],
\end{equation}

where $E$ and $l$ represent energy and angular momentum constants, respectively. It
has been found more practical to introduce a new variable $u$, such that, $u=%
\frac{1}{r}.$ \ Using this transformation, Eq.(5) leads to
\begin{equation}
\frac{d^{2}u}{d\varphi ^{2}}=2u^{3}\kappa(u)+\frac{u^{4}}{2}\frac{d\kappa(u)%
}{du},
\end{equation}%
in which $\kappa(u)=\frac{K(r)}{h(r)\left[g(r)E+f(r)l)\right]^{2}}\left[p(r)E^{2}-f(r)l^{2}-2g(r)lE\right]$. The solution of Eq.(6) will be
used to define another equation in the following way%
\begin{equation}
A(r,\varphi )\equiv \frac{dr}{d\varphi }.
\end{equation}%
If the direction of the orbit is denoted by $d$ and that of the
coordinate line $\varphi =$ constant $\delta ,$ we have%
\begin{eqnarray}
d &=&\left( dr,d\varphi \right) =\left( A,1\right) d\varphi \text{ \ \ \ \ \
\ \ }d\varphi <0,  \notag \\
\delta &=&\left( \delta r,0\right) =\left( 1,0\right) \delta r.
\end{eqnarray}%
Using these definitions in (2), we get

\begin{equation}
\tan \left( \psi \right) =\frac{\left[ h^{-1}(r)p(r)\right] ^{1/2}}{%
\left\vert A(r,\varphi )\right\vert }.
\end{equation}%
Then, the one-sided bending angle is defined as $\epsilon =\psi
-\varphi .$

\section{EXAMPLES of GRAVITATIONAL LENSING in ROTATING GEOMETRIES}
\subsection{The Rotating Bertotti - Robinson Space-Time }
The Bertotti - Robinson  space-time is the sole conformally flat solution of the Einstein - Maxwell equations, which describes the universe filled with uniform non-null electromagnetic field. It is non-null as $ F_{\mu\nu}F^{\mu\nu}\neq 0 $ and uniform because the null tetrad component of the Ricci tensor $\Phi_{11}$ is constant. The rotating version of the BR solution was obtained by Carter \cite{3}, and the related solution is described by the metric \cite{2}

\begin{equation}
ds^{2}=\frac{F( \theta)}{r^{2}} \left [dt^{2}-dr^{2}-r^{2}d\theta ^{2}-\frac{%
r^{2}\sin ^{2}\theta}{F^{2}(\theta)}\left ( d\varphi - \frac{q}{r} dt \right
)^{2} \right ] ,
\end{equation}

where $a$ is the constant rotation parameter, $F(\theta)=1+a^{2}(1+cos^{2}(\theta))$
and $q=2a\sqrt{1+a^{2}}$.

The physical description of this solution is achieved via the Newman - Penrose (NP) formalism. The set of proper null tetrads $1-form$ is given by

\begin{equation}
\begin{gathered} l=\frac{\sqrt{F(\theta)}}{2\sqrt{2}r}(dt-dr),\\
n=\frac{2\sqrt{F(\theta)}}{\sqrt{2}r}(dt+dr),\\
m=\frac{i\sqrt{F(\theta)}}{\sqrt{2}}+\frac{sin(\theta)}{\sqrt{2F(\theta)}}%
\left(\frac{2a}{r}\sqrt{1+a^{2}}dt-d\varphi\right). \end{gathered}
\end{equation}

The non-zero Weyl and Ricci scalars are

\begin{equation}
\begin{gathered} \Psi_{2}
=\frac{a^{2}}{F^{3}(\theta)}\left[(1+a^{2})cos(2\theta)+a^{2}cos^{2}(%
\theta)-\frac{i}{a}\sqrt{1+a^{2}}cos(\theta)\left(1+2a^{2}+a^{2}sin^{2}(%
\theta)\right)\right],\\ \Phi_{11}=\frac{1}{4F^{2}(\theta)}. \end{gathered}
\end{equation}

The contribution of rotation parameter $ a $ to the Weyl and Ricci scalars indicates that the resulting space-time character is of  Petrov type - D. Furthermore, the rotation parameter breaks the uniform nature of the em field. This non-uniformity in the em field creates an anisotropy when compared to the non-rotating BR solution. The change in the em energy density of anisotropy between $\theta=0^{o}$ and $ \theta=1^{o}$  directions can be found by
\begin{equation}
\Delta\Phi_{11}=\Phi_{11}(\theta=1^{o})-\Phi_{11}(\theta=0^{o})=\frac{%
0.0025a^{2}}{(1+1.99a^{2})(1+2a^{2})}.
\end{equation}

The variation in the em energy density is plotted in Fig.2, against the rotation paramater $ a $. From (12), we can easily compute the maximum anisotropy by comparing the distribution of energy density along $\theta=0$ and $\theta=\pi/2$. By doing so, we obtain

\begin{equation}
\frac{\Phi_{11}(\theta=0)}{\Phi_{11}(\theta=\pi/2)}=\left( \frac{1+a^{2}}{1+2a^{2}} \right)^{2} ,
\end{equation}

which suggests that for fast rotations $(a \rightarrow \infty)$, the maximum anisotropy is of the order $1/4$. Stated otherwise, if our universe is governed entirely by rotating em radiation, the maximum possible distortion along the $z-$axis is $1/4$ times the distortion along the equatorial direction. This is not unusual, as the rotational effects create bulges in the $\theta=\pi/2$ plane compared with the $\theta=0$ direction in analogy with the flattening of the rotating Earth.

\begin{figure}[htp]
\par
\label{figure}
\par
\begin{tabular}{cc}
\includegraphics[width=80mm]{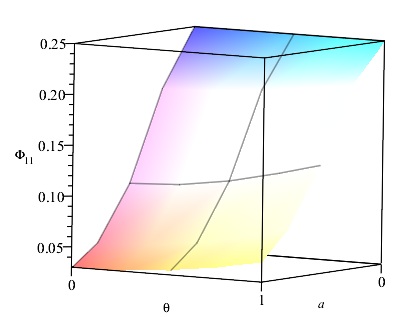} &
\end{tabular}
\caption{At $a=0$ we have complete isotropy and maximum energy density $\Phi_{11}=1/4$. For increasing $a$, $\Phi_{11}$ decreases for $a\rightarrow \infty$. Angular dependence is shown for a very narrow angle of $\Delta\Phi_{11}$.    }\centering
\end{figure}

This anisotropy in the electromagnetic field influences the light rays that propagates within it. To show this effect, we calculate the bending angle of light, by using the method of RI.

By using Eq.(6), we get
\begin{equation}
\frac{d^{2}u}{d\varphi ^{2}}-\left[\frac{qE^{3}}{(1+a^{2})^{3}\gamma^{3}(u)} %
\right]u=-\left[ \frac{l}{(1+a^{2})\gamma(u)}+\frac{4aEl}{%
(1+a^{2})^{3/2}\gamma^{3}(u)} \right]u^{2}-\left[\frac{%
qE(1+l^{2}-3a^{2}l^{2})}{(1+a^{2})^{2}\gamma^{3}(u)}+\frac{1}{%
(1+a^{2})\gamma^{2}(u)} \right]u^{3},
\end{equation}

where $\gamma(u)=\frac{qE}{1+a^{2}}+\frac{l}{1+a^{2}}%
\left((1+a^{2})^{2}-q^{2}\right)u.$

Since $u<<1$, Eq.(15) simplifies to

\begin{equation}
\frac{d^{2}u}{d\varphi ^{2}}-\omega^{2} u=-\mu u^{2} -\nu u^{3},
\end{equation}

where $\omega=\frac{1}{q}$, $\mu=\frac{l}{qE}+\frac{4al(1+a^{2})^{3/2}}{q^{3}}
$ and $\nu=\frac{(1+a^{2})(2+l^{2}-3a^{2}l^{2})}{E^{2}q^{2}}$.

The first approximate solution, $u=\frac{e^{\omega \varphi }}{R},$ is the solution of the homogeneous part of Eq.(16). This solution corresponds to the undeflected line. If this solution is substituted back in Eq.(16), the perturbed solution for $u$ is obtained as

\begin{equation}
u(\varphi)=\frac{e^{\omega \varphi}}{R}+\frac{\nu}{8\omega^{2}R^{3}}%
e^{3\omega\varphi}+\frac{\mu}{3\omega^{2}R^{2}}e^{2\omega\varphi},
\end{equation}
and the Eq.(7) becomes
\begin{equation}
A(r,\varphi)=-r^{2}\left[ \frac{\omega}{R}e^{\omega \varphi}+\frac{3\nu}{%
8\omega R^{3}}e^{3\omega\varphi}+\frac{2\mu}{3\omega R^{2}}%
e^{2\omega\varphi} \right].
\end{equation}

Here $R$ is a constant parameter related to the physically meaningful area distance $r_{0}$\ of the closest approach calculated at
 $\varphi=\pi /2$
which is found to be%
\begin{equation}
\frac{1}{r_{0}}=\frac{e^{\omega \pi/2}}{R}-\frac{\nu}{8\omega^{2}R^{3}}%
e^{3\omega\pi/2}-\frac{\mu}{3\omega^{2}R^{2}}e^{\omega\pi}.
\end{equation}

We calculate the bending angle when $\varphi =0$ that corresponds to a large distance from the source. This small angle approximation leads us to set $\tan\psi \approx \psi_{0}$. For this particular case, we found \\
\begin{equation}
r\approx R,\text{ \ \ \ \ \ } A(r,\varphi=0)\approx-\omega R,
\end{equation}

so that, the one - sided bending angle becomes
\begin{equation}
\epsilon=\psi_{0}=\frac{1}{\omega R^{2}}=\frac{q}{R^{2}}=\frac{2a\sqrt{
1+a^{2}}}{R^{2}}.
\end{equation}

The effect of rotation on the bending angle of light is depicted in Fig.3.

\begin{figure}[htp]
\par
\label{figure}
\par
\begin{tabular}{cc}
\includegraphics[width=100mm]{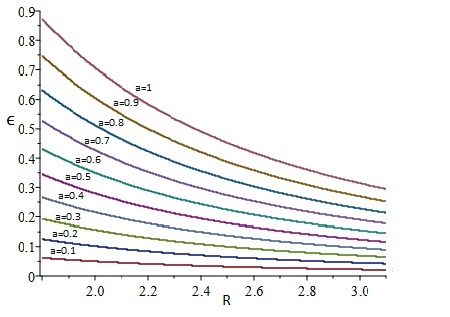} &
\end{tabular}
\caption{Variation of the bending angle both as a function of rotation parameter $a$ and the distance $R$. Evidently, rotation of the universe increases the bending angle, which is expected.}\centering
\end{figure}

\subsection{Rotating Axion - coupled Electromagnetic Field}
The singularity-free colliding gravitational wave solution coupled with axion field was given in \cite{10}. The interaction region represents the solution of Einstein-Maxwell-dilaton-axion equations in the limit of zero dilaton field. The obtained solution reduces to the Bell-Szekeres (BS) solution, which describes the non-linear interaction of shock em waves in the absence of axion field. The remarkable feature of the BS solution is that the interaction region is locally isometric to the part of the BR solution. The resulting metric that describes the collision of em field coupled with axion is given by \\

\begin{equation}
ds^{2}=2dudv-\Delta dy^{2}-\delta (dx+q_{0} \tau dy)^2,
\end{equation}
in which the used notations represent
\begin{equation}
\Delta = 1-\tau ^{2}, \text{ \ \ \ } \delta = 1-\sigma^{2}, \text{ \ \ \ } \tau = \sin (au\theta(u)+bv\theta(v)),  \\
 \sigma = \sin (au\theta(u)-bv\theta(v))   \text{ \ \ \ } and \text{ \ \ \ } q_{0} = constant .
\end{equation}
Here, $(u,v)$ are the double null coordinates, $(a,b)$ are the constant electromagnetic parameters,  $ (\theta(u), \theta(v))$ are the unit step functions and $(\tau, \sigma)$ are the prolate coordinates. The constant parameter $q_{0}$ is related to the axion field and the physically acceptable solution to the field equations are obtained when $q_{0}=1$. If we set $q_{0}=0$, the axion vanishes and the solution reduces to BS solution. The metric in $(\tau, \sigma, x, y)$ coordinates becomes

\begin{equation}
ds^{2}=\frac{1}{2ab}(\frac{d\tau^{2}}{\Delta}-\frac{d\sigma^{2}}{\delta})-\Delta dy^{2}-\delta (dx+ \tau dy)^2.
\end{equation}

The metric in the usual $(t,r,\theta,\varphi)$ coordinates is obtained via the following transformations
\begin{equation}
\tau=\frac{1}{2r}(r^{2}-t^{2}+1), \text{ \ \ \ } \sigma=\cos\theta, \text{ \ \ \ } \tanh y = \frac{1}{2t}(r^{2}-t^{2}-1), \text{ \ \ \ } x=\varphi-\frac{1}{2}\ln\mid \frac{(r+t)^{2}-1}{(r-t)^{2}-1} \mid ,
\end{equation}

from which we have removed the overall constant factor $1/2ab$ by metric scaling. This transformation leads us to express the resulting metric in the BR form as

\begin{equation}
ds^{2}=\frac{1}{r^{2}} \left [dt^{2}-dr^{2}-r^{2}d\theta ^{2}-r^{2}\sin
^{2}\theta\left ( d\varphi - \frac{1}{r} dt \right )^{2} \right ] .
\end{equation}

This metric represents a space-time filled with em field coupled to an axion field.
Next, we calculate the contribution of a mixture of these fields to the bending angle of light. As before, we start our calculation with Eq.(6),
which yields
\begin{equation}
\frac{d^{2}u}{d\varphi ^{2}}-u=-\alpha u^{2}-\beta u^{3}.
\end{equation}

where $\alpha=\frac{l}{E}$ and $\beta=\frac{(2+l^{2})}{E^{2}}$.

The homogeneous part of Eq.(27) has a solution of the form $u=\frac{e^{ \varphi }}{R}$,  This solution is substituted
back into Eq.(27) and the perturbed resulting solution for $u$ is obtained as

\begin{equation}
u(\varphi)=\frac{e^{ \varphi}}{R}+\frac{\beta}{8R^{3}}e^{3\varphi}+\frac{%
\alpha}{3R^{2}}e^{2\varphi},
\end{equation}
and the Eq.(7) becomes
\begin{equation}
A(r,\varphi)=-r^{2}\left[ \frac{1}{R}e^{\varphi}+\frac{3\beta}{8 R^{3}}%
e^{3\varphi}+\frac{2\alpha}{3 R^{2}}e^{2\varphi} \right].
\end{equation}

The closest approach distance $r_{0}$\ is calculated at $\varphi =\pi /2,$
and is given by%
\begin{equation}
\frac{1}{r_{0}}=\frac{e^{ \pi/2}}{R}+\frac{\beta}{8R^{3}}e^{3\pi/2}+\frac{\alpha}{%
3R^{2}}e^{\pi}.
\end{equation}

We calculate the bending angle when $\varphi =0$, which is the bending
angle measured for a long distance from the source. For this particular
case we have
\begin{equation}
r\approx R,\text{ \ \ \ \ \ } A(r,\varphi=0)\approx- R,
\end{equation}

hence, the one - sided bending angle becomes
\begin{equation}
\epsilon=\psi_{0}=\frac{1}{ R^{2}}.
\end{equation}
It is important to note that the bending  arises due to the existence of the axion field that creates rotation in the universe. Since we have chosen $q_{0}=1$, its presence in the calculated bending angle is not apparent.

\subsection{Lensing in the Twisting NUT Universe}

The NUT metric is written as

\begin{equation}
ds^{2}=f(r)\left[ dt+2lcos\theta d\varphi^{2} \right]^{2}-\frac{dr^{2}}{f(r)}%
-\left( r^{2}+l^{2} \right)\left[d\theta^{2}+sin^{2}\theta d\varphi^{2}%
\right],
\end{equation}

where $f(r)=1-\frac{2(mr+l^{2})}{r^{2}+l^{2}}.$

This is another vacuum metric that generalizes Schwarzschild metric in analogy with the Kerr metric. The NUT parameter is specified by $l$ and the metric reduces to Schwarzschild for $l=0$. For the physical interpretation of NUT metric, there are different views \cite{12,13}. We adapt the interpretation that $l$ corresponds to the twist of the vacuum space-time \cite{12}. We arrive at this interpretation by consideration of a general class of Einstein-Maxwell solutions in the limit when the em field vanishes \cite{13}. The space-time left as a result corresponds exactly to an isometric spacetime of the NUT solution.

The source-free em field gives, in an appropriate NP tetrad \cite{12},

\begin{equation}
\Psi_{2}=\frac{-mp}{r^{6}}\left \{ r^{3}+3m(p-1)(r-2m)r+2m^{3}\frac{(p-1)^{2}%
}{p}+i(p^{2}-1)^{1/2}\left[(3m-r)r^{2}-2m^{2}\frac{(p-1)}{p}\right]\right\}
\end{equation}

and the NUT parameter

\begin{equation}
l=\pm m \sqrt{p-\frac{1}{p}},
\end{equation}

in which $p$ is the twist parameter left-over from the twisting spacetime. Using
Eq.(6), we get
\begin{equation}
\frac{d^{2}u}{d\varphi ^{2}}=\frac{2ul^{2}}{h^{2}}+\frac{2u^{3}l^{4}}{h^{2}}%
-(u+2l^{2}u^{3})f(u)-\left[\frac{u^{2}+l^{2}u^{4}}{2}\right]\frac{df(u)}{du},
\end{equation}

where $f(u)=1-\frac{2u^{2}(\frac{m}{u}+l^{2})}{1+u^{2}l^{2}}$ and $h$ is the constant measuring the angular momentum of the test particle.

Since $u<<1$, Eq.(36) simplifies to

\begin{equation}
\frac{d^{2}u}{d\varphi ^{2}}+\delta^{2}u=3mu^{2}+\rho
u^{3}+5ml^{2}u^{4}+6l^{4}u^{5},
\end{equation}

where $\delta^{2}=1-\frac{2l^{2}}{h^{2}}$ and $\rho=\frac{2l^{4}}{h^{2}}%
+2l^{2}$. We have two different cases.

\subsubsection{ The case of $1>\frac{2l^{2}}{h^{2}}$ (small NUT parameter):}

The first approximate solution, $u=\frac{sin(\delta\varphi)}{R},$ is
substituted back in Eq.(37) and its resulting solution for $u$ is obtained as

\begin{equation}
\begin{aligned}
u(\varphi)=\frac{sin(\delta\varphi)}{R}+\frac{1}{24\delta^{2}R^{5}}&\left%
\{(-6l^{4}sin(\delta\varphi)-8l^{2}mR)cos^{4}(\delta\varphi)+cos^{2}(\delta%
\varphi)[(3\rho
R^{2}+27l^{4})sin(\delta\varphi)+48l^{2}mR+24mR^{3}]\right.\\
&\left.-9\delta\varphi(\rho
R^{2}+5l^{4})cos(\delta\varphi)+sin(\delta\varphi)(6\rho
R^{2}+24l^{4})+24mR(l^{2}+R^{2}) \right\}, \end{aligned}
\end{equation}

so that Eq.(7) becomes

\begin{equation}
\begin{aligned} A(r,\varphi)&=-r^{2}\left\{\frac{\delta
cos(\delta\varphi)}{R}+\frac{1}{24\delta^{2}R^{5}}\left\{-6l^{4}\delta
cos^{5}(\delta\varphi)-4\delta
cos^{3}(\delta\varphi)sin(\delta\varphi)(-6l^{4}sin(\delta%
\varphi)-8l^{2}mR)+\delta cos^{3}(\delta\varphi)(3\rho R^{2}+27l^{4})
\right. \right. \\ &\left. \left. -\delta sin(2\delta\varphi)[(3\rho
R^{2}+27l^{4})sin(\delta\varphi)+48l^{2}mR+24mR^{3}]-9\delta
cos(\delta\varphi)( \rho R^{2}+5l^{4}) +9\delta^{2}\varphi
sin(\delta\varphi)(\rho R^{2}+5l^{4}) \right. \right. \\ &\left.
\left.+\delta cos(\delta\varphi)(6\rho R^{2}+24l^{4}) \right\} \right\}.
\end{aligned}
\end{equation}

The closest approach distance $r_{0}$\ is calculated at $\varphi =\pi /2,$
which is found to be
\begin{equation}
\frac{1}{r_{0}}=\frac{1}{R}+\frac{m}{\delta^{2}R^{2}}+\frac{m\rho^{2}}{%
4\delta^{2}R^{3}}+\frac{ml^{2}}{\delta^{2}R^{4}}+\frac{l^{2}}{\delta^{2}R^{5}%
}.
\end{equation}

We calculate the bending angle when $\varphi =0$ which is the bending
angle named as the small angle $\psi_{0}$ and $R>>1$. For this particular
case we found that
\begin{equation}
r\approx \frac{\delta^{2}R^{2}}{2m},\text{ \ \ \ \ \ } A(r,\varphi=0)%
\approx- r^{2}\frac{\delta}{R},
\end{equation}

and the one-sided bending angle is
\begin{equation}
\epsilon =\psi _{0}\simeq \frac{2m}{\delta^{3}R}\left\{ 1-\frac{4m^{2}}{%
\delta^{2}R^{2}}+\frac{ 8m^{2}l^{2}}{\delta^{4}R^{4}}\right\}^{1/2} \simeq%
\frac{2m}{\delta^{3}R}\left\{ 1-\frac{2m^{2}}{\delta^{2}R^{2}}+\frac{
4m^{2}l^{2}}{\delta^{4}R^{4}}\right\} +\mathcal{O}\left( \frac{m^{5}l ^{4}}{%
R^{9}}\right) .
\end{equation}

This particular case implies that the effect of NUT parameter is very small. Hence, the mass term dominates all the others and the obtained bending angle becomes similar to the Schwarzchild case, which is convex lensing.

\subsubsection{ The case of $1<\frac{2l^{2}}{h^{2}}$ (large NUT parameter):}

In this case Eq.(36) takes the from

\begin{equation}
\frac{d^{2}u}{d\varphi ^{2}}-\zeta^{2}u=3mu^{2}+\rho
u^{3}+5ml^{2}u^{4}+6l^{4}u^{5},
\end{equation}

where $\zeta^{2}=\frac{2l^{2}}{h^{2}}-1$.

The first approximate solution, $u=\frac{e^{\zeta \varphi }}{R},$ is
substituted back in Eq.(43) and its resulting solution for $u$ is obtained as

\begin{equation}
u(\varphi)=\frac{e^{\zeta \varphi}}{R}-\frac{l^{4}}{\zeta^{2}R^{5}}%
e^{5\zeta\varphi}-\frac{5ml^{2}}{\zeta^{2}R^{4}}e^{4\zeta\varphi}-\frac{\rho%
}{\zeta^{2}R^{3}}e^{3\zeta\varphi}-\frac{3m}{\zeta^{2}R^{2}}%
e^{2\zeta\varphi},
\end{equation}
and Eq.(7) becomes
\begin{equation}
A(r,\varphi)=-r^{2}\left[ \frac{\zeta e^{\zeta \varphi}}{R}-\frac{5l^{4}}{%
\zeta R^{5}}e^{5\zeta\varphi}-\frac{20ml^{2}}{\zeta R^{4}}e^{4\zeta\varphi}-%
\frac{3\rho}{\zeta R^{3}}e^{3\zeta\varphi}-\frac{6m}{\zeta R^{2}}%
e^{2\zeta\varphi} \right].
\end{equation}

The closest approach distance $r_{0}$\ is calculated at $\varphi =\pi /2,$
which is found to be
\begin{equation}
\frac{1}{r_{0}}=\frac{e^{\zeta \pi /2}}{R}-\frac{l^{4}}{\zeta^{2}R^{5}}%
e^{5\zeta\pi /2}-\frac{5ml^{2}}{\zeta^{2}R^{4}}e^{2\zeta\pi}-\frac{\rho}{%
\zeta^{2}R^{3}}e^{3\zeta\pi /2}-\frac{3m}{\zeta^{2}R^{2}}e^{\zeta\pi}.
\end{equation}

We calculate the bending angle when $\varphi =0$ , which yields for this particular
case,
\begin{equation}
r\approx R,\text{ \ \ \ \ \ } A(r,\varphi=0)\approx- \zeta R,
\end{equation}

so that the one-sided bending angle is
\begin{equation}
\epsilon =\psi _{0}\simeq \frac{1}{\zeta R}\left\{ 1-\frac{2m}{R^{2}}+\frac{
2l^{2}}{R^{4}}\right\}^{1/2}\simeq\frac{1}{\zeta R}\left\{ 1-\frac{m}{R^{2}}+%
\frac{ l^{2}}{R^{4}}\right\}+\mathcal{O}\left( \frac{l ^{4}}{R^{9}}\right) .
\end{equation}

This is the case where the NUT parameter dominates the mass term. The calculated  bending angle indicates that there is a drastic decrease in the bending angle. The reason of this is the concave role played by the NUT parameter on the geometry.

\section{Results and Discussions}

The gravitational lensing in rotating and twisting universes
are studied. The considered model of universes are assumed to develop whenever the gravitational waves coupled with em and axion waves interact nonlinearly and give rise to high curvature zone in the fabric of space-time. Hence, high curvature zone space-time structures constitute another mechanism that cause the light to bend. Moreover, the rotation and twisting parameters are the key parameters that take part in gravitational lensing. Let us add that in this study we did not consider the case of extreme curvature zones, as it is in the inner horizon of a black hole where the light bending creates photon spheres.\\

A striking result is obtained in the NUT space-time. The solution is dependent on two
different behaviour of the NUT parameter; thus, we investiged two different
bending angles based on the mathematically bounded NUT parameter to angular momentum ratio. We noticed that when the NUT parameter is small, the bending angle comes out to be almost the same as the one for the Schwarzchild case, which is convex lensing. On the other hand, the large NUT parameter generates a diverging (concave) lens effect. We recall that $90 \%$ of the universe consists of voids due to dark energy. During the first ten billion years following the Big Bang, dark matter used to dominate and as a result bending of light was convex type. Later on, the mysterious dark energy took over and  we have in the present era an accelerating expansion of universe which implies concave lensing from the hyperbolic nature of the space-time. Note that in this description we exclude the local  sources that might create regional convex lensing. Our conclusion is that concave nature of the light bending becomes inevitable in an expanding universe of voids.   \\

To sum up, we studied the gravitational lens effect of three different non-singular cosmological models that results as a nonlinear interaction of gravitational waves coupled with em and axion waves. Calculations have revealed that not only the gravitational field of massive objects is effective on the bending angle of light, at the same time, the high curvature zone of the space-time is also effective. It remains to be seen whether deflection of light by voids in absence of massive objects can be considered as an indirect evidence of the mysterious dark energy. Finally let us add that whatever has been done$/$said for twisting$/$rotating vacua of space-time are valid also for the mysterious dark matter since its gravitational effect cannot be suppressed.


\begin{thebibliography}{99}
\bibitem{1} M. Halilsoy, S. H. Mazharimousavi and O. Gurtug, JCAP \textbf{11}, 010, (2014).
\bibitem{2} M. Halilsoy and O. Gurtug, Phys. Rev. \textbf{D} \textbf{75}, 124021, (2007).
\bibitem{3} B. Carter, Black Hole Equilibrium States, in Black Holes, entitled by C.M Dewitt (Gordon and Breach, New York,1973).

\bibitem{4} A. Al-Badawi and M. Halilsoy, Il Nuovo Cimento, \textbf{119}, 931,  (2004).

\bibitem{5} M. Halilsoy, Phys. Rev. \textbf{D} \textbf{37}, 2121, (1988).


\bibitem{6} M. Halilsoy, Jour. Math. Phys., \textbf{31}, 2694, (1990).

\bibitem{7} P. Bell and P. Szekeres, Gen. Relativ. Gravit., \textbf{5}, 275, 1974.

\bibitem{8} J. B. Griffith, Colliding Plane Waves in General Relativity (Oxford University Press, 1991).

\bibitem{9} E. Newman and R. Penrose, Jour. Math. Phys., \textbf{3}, 566, (1962).

\bibitem{10} M. Halilsoy and I. Sakalli, Class. Quantum Grav., \textbf{20},1417, (2003).

\bibitem{11} E. T. Newman, L. Tamburino and T. Unti, Jour. Math. Phys., \textbf{4}, 915, (1963).


\bibitem{12} A. Al-Badawi and M. Halilsoy, Gen. Relativ. Gravit., \textbf{38}, 1729, (2006).

\bibitem{13} D. Lynden-Bell and M. Nouri-Zonoz, Rev. Mod. Phys., \textbf{70}, 427, (1998).

\bibitem{14} H. He and Z. Zhang, JCAP, \textbf{08}, 036, (2017).

\bibitem{15} F. Dong, J. Zhang, Y. Yu, X. Yang, H. Li, J. Han, W. Luo, J. Zhang and L. Fu, ApJ,
\textbf{874}, 7 (2019).

\bibitem{16} W. Rindler and M. Ishak, Phys. Rev. \textbf{D}
\textbf{76}, 043006 (2007).

\bibitem{17} O. Gurtug and M. Mangut, Phys. Rev. \textbf{D} \textbf{99}, 084003, (2019).

\bibitem{18} J. Sultana and D. Kazanas, Phys. Rev. \textbf{D} \textbf{81}, 125502, (2010).

\bibitem{19} M. Ishak, W. Rindler, J. Dossett, J. Moldenhauer and C. Allison, Mon. Not. Roy. Astron. Soc.  \textbf{335}, 1279, (2008).

\bibitem{20} O. Gurtug and M. Mangut,  Annalen der Physik (Berlin), 1900576, (2020).

\end{thebibliography}
\end{document}